\documentclass[12pt]{article}
\usepackage[dvips]{color}
\usepackage{amsmath}
\usepackage{amsfonts}
\usepackage{graphicx}
\usepackage{amssymb}

\begin{document}
\title{\bf Lagrangian Formulation of an Infinite Derivative Real Scalar Field Theory in the Framework of the Covariant Kempf-Mangano Algebra in a $\bf(D+1)$-dimensional Minkowski Space-time}

\author{A. Izadi, S. K. Moayedi \thanks{Corresponding author, E-mail:
s-moayedi@araku.ac.ir}\hspace{1mm}\\
{\small {\em  Department of Physics, Faculty of Sciences,
Arak University, Arak 38156-8-8349, Iran}}\\
}
\date{\small{}}
\maketitle
\begin{abstract}
\noindent In 2017, G. P. de Brito and co-workers suggested a covariant generalization of the Kempf-Mangano algebra in a $(D+1)$-dimensional Minkowski space-time [A. Kempf and G. Mangano, Phys. Rev. D \textbf{55}, 7909 (1997); G. P. de Brito, P. I. C. Caneda, Y. M. P. Gomes, J. T. Guaitolini
Junior, and  V. Nikoofard, Adv. High Energy Phys. \textbf{2017}, 4768341 (2017)]. It is shown that reformulation of a real scalar field theory from the viewpoint of the covariant  Kempf-Mangano algebra leads to an infinite derivative Klein-Gordon wave equation which describes two bosonic particles in the free space (a usual particle and a ghostlike particle). We show that in the low-energy (large-distance) limit our infinite derivative scalar field theory behaves like a Pais-Uhlenbeck oscillator for a spatially
homogeneous field configuration $\phi(t,\vec{\textbf{x}})=\phi(t)$. Our calculations show that there is a characteristic length scale $\delta$ in our model whose upper limit in a four-dimensional Minkowski space-time is close to the nuclear scale, i.e., $\delta_{max}\sim \delta_{nuclear\ scale}\sim 10^{-15}\, m$. Finally, we show that there is an equivalence between a non-local real scalar field theory with a non-local form factor ${\cal K}(x-y)= -\frac{\square_x}{(1-\frac{\delta^2}{2}\square_x)^2} \ \delta^{(D+1)}(x-y)$ and an infinite derivative real scalar field theory from the viewpoint of the covariant Kempf-Mangano algebra.

\noindent
\hspace{0.35cm}



\end{abstract}

\section{Introduction}
In classical mechanics, the motion of a point particle is described
by the following action functional
\begin{equation*}
S[x]= \int_{t_i}^{t_f} dt \ L(x(t),\dot{x}(t)).
\end{equation*}
In 1850, the Russian mathematician Mikhail Ostrogradsky proposed a
higher-order generalization of Lagrangian mechanics [1]. The action
functional for a higher-order derivative mechanical system is
\begin{equation*}
S[x]= \int_{t_i}^{t_f} dt \
L(x^{(0)}(t),x^{(1)}(t),\ldots,x^{(N)}(t)),
\end{equation*}
where $x^{(i)}(t):=\frac{d^{i}x(t)}{dt^{i}}$ and $i\in
\{0,1,\ldots,N\}$.\\
The variation of the above action functional with respect to $x$
leads to the following generalized Euler-Lagrange equation
\begin{equation}
\sum_{i=0}^{N} \left(- \frac{d}{dt}\right)^{i} \frac{\partial
L}{\partial x^{(i)}}=0.
\end{equation}
In 1942, B. Podolsky suggested a higher-order derivative
generalization of Maxwell electrodynamics, in which the
electrostatic self-energy of a point charge was a finite value
[2].\\
Six years later, A. E. S. Green presented a higher-order derivative
meson-field theory, in which the potential energy for a point
nucleon at the origin was a finite value [3]. In 1950, A. Pais and
G. E. Uhlenbeck showed that the appearance of higher-order
derivative terms in the Lagrangian density of a quantum field theory
can eliminate the ultraviolet divergences that appear in the
$S$-matrix elements [4].\\
Today we know that the addition of higher-order derivative terms to
the action functional of a quantum field theory is a possible way of
regularizing quantum field theories [5-9]. Recently, the infinite
derivative scalar field theory of the form
\begin{equation}
S[\phi] =\underbrace{\frac{1}{2c} \int d^{D+1} x \ d^{D+1} y \
\phi(x) \ {\cal K}(x-y)\ \phi(y)}_{kinetic \ term } \\
-\underbrace{\frac{1}{c} \int d^{D+1} x \ V(\phi(x))}_{local \
potential \ term}+\underbrace{\frac{1}{c} \int d^{D+1} x \ J(x) \
\phi(x)}_{interaction \ term},
\end{equation}
has attracted a considerable attention because of its importance in
non-local quantum field theory and string field theory [10].\\
The operator ${\cal K}(x-y)$ in Eq. (2) has the following general
form [10]:
\begin{equation*}
{\cal K}(x-y)= F(\square_x) \ \delta^{(D+1)}(x-y),
\end{equation*}
where $F(\square_x)$ is an entire analytic function of the
d'Alembertian operator $\square_x$. On the other hand, different
theories of quantum gravity such as string theory and loop quantum
gravity predict the existence of a minimal length scale of the order
of the Planck length [11-15].  Today we know that the existence of a
minimal length scale in different theories of quantum gravity leads
to a generalization of Heisenberg uncertainty principle as follows
[12]:
\begin{equation*}
\Delta X \Delta P \geqslant {\hbar \over 2} \left[1+ \beta (\Delta
P)^2+\cdots\right], \ \ \ \beta>0.
\end{equation*}
The above generalized uncertainty principle leads to a minimal
length scale $\Delta X_0 = \hbar\sqrt{\beta} \ (\Delta X_0>0)$ in
the measurement of space intervals [12]. It should be noted that the
reformulation of quantum field theory in the presence of a minimal
length scale is another possible way for regularizing a quantum
field theory [12]. The aim of this paper is reformulation of a real
scalar field theory from the viewpoint of a covariant generalization
of the Kempf-Mangano algebra which was proposed by G. P. de Brito
and co-workers in Ref. [14].\\This paper is organized as follows. In
Section 2, a covariant generalization of the Kempf-Mangano algebra
in a $(D+1)$-dimensional Minkowski space-time is presented briefly.
In Section 3, we show that reformulation of a real scalar field
theory in the framework of the covariant Kempf-Mangano algebra leads
to a generalized Klein-Gordon wave equation with infinitely many
derivatives. The free space solutions of this generalized
Klein-Gordon wave equation describe two bosonic particles. In
Section 4, we show that in the low-energy (large-distance) limit the
infinite derivative scalar field theory which was formulated in
Section 3 behaves like a Pais-Uhlenbeck oscillator for a spatially
homogeneous field configuration $\phi(t,\vec{\textbf{x}})=\phi(t)$
[4].\\
Our calculations in Sections 3 and 4 together with numerical
evaluations in Section 5 show that there is a characteristic length
scale in our generalized real scalar field theory whose upper limit
is very near to the nuclear scale, i.e., $10^{-15}\ m$. Finally, it
should be emphasized that the results of this paper in the
low-energy regime are compatible with the results of the standard
Klein-Gordon theory. We use SI units in this paper. The flat
space-time metric has the signature
$\eta_{\mu\nu}=\eta^{\mu\nu}=diag(+, \underbrace{-,\ldots,-}_{D \
times}).$

\section{The Covariant Kempf-Mangano Algebra}
In 1997, A. Kempf together with G. Mangano proposed a one-parameter
extension of the Heisenberg algebra [12]. The Kempf-Mangano algebra
in a $D$-dimensional Euclidean space is described by the following
generalized commutation relations:
\begin{eqnarray}
\nonumber\left[\hat{X}^{i},\hat{P}^{j}\right] &=& i\hbar \left(
\frac{\ell^{2}\hat{\vec{\textbf{P}}}^{2}}{(1+2\ell^2 \hat{\vec{\textbf{P}}}^{2})^\frac{1}{2}-1}\delta^{ij}+\ell^2 \hat{P}^i \hat{P}^j\right) \\
&=& i\hbar \left(\frac{1+(1+2\ell^2 \hat{\vec{\textbf{P}}}^{2})^\frac{1}{2}}{2}\delta^{ij}+\ell^2 \hat{P}^i \hat{P}^j\right), \\
\left[\hat{X}^{i},\hat{X}^{j}\right] &=& 0, \\
\left[\hat{P}^{i},\hat{P}^{j}\right] &=& 0,
\end{eqnarray}
where $i,j=1,2,\ldots,D$ and $\ell$ is a non-negative constant
parameter with dimension of $[momentum]^{-1}$ [12]. The
reformulation of non-relativistic quantum mechanics and a charged
scalar field in the framework of Kempf-Mangano algebra have been
studied in details  in Ref. [13].\\
In 2017, G. P. de Brito and co-workers proposed a covariant
generalization of the Kempf-Mangano algebra [14]. The covariant
Kempf-Mangano algebra in a $(D+1)$-dimensional Minkowski space-time
is described by the following generalized commutation
relations:\footnote{In 2006, C. Quesne and V. M. Tkachuk introduced
a two-parameter extension of the covariant Heisenberg algebra in a
$(D+1)$-dimensional space-time [15]. There are many papers about
reformulation of quantum field theory from the viewpoint of the
Quesne-Tkachuk algebra. For a review, we refer the reader to Refs.
[16,17].}
\begin{eqnarray}
\left[\hat{X}^{\mu},\hat{P}^{\nu}\right] &=& -i\hbar \left(\frac{1+(1-2\ell^2 \hat{\textbf{P}}^{2})^\frac{1}{2}}{2}\eta^{\mu\nu}-\ell^2 \hat{P}^{\mu} \hat{P}^{\nu}\right), \\
\left[\hat{X}^{\mu},\hat{X}^{\nu}\right] &=& 0, \\
\left[\hat{P}^{\mu},\hat{P}^{\nu}\right] &=& 0,
\end{eqnarray}
Where $\mu,\nu=0,1,2,\ldots,D$, $\hat{X}^{\mu}$ and $\hat{P}^{\mu}$
are the generalized position and momentum operators, and
$\hat{\textbf{P}}^{2}=\hat{P}_{\mu}\hat{P}^{\mu}=(\hat{P}_{0})^{2}-\sum_{i=1}^{D}(\hat{P}^{i})^{2}$.
In the coordinate representation, the generalized position and
momentum operators $\hat{X}^{\mu}$ and $\hat{P}^{\mu}$ in Eqs.
(6)-(8) have the following exact representations [14]:
\begin{eqnarray}
\hat{X}^{\mu} &=& \hat{x}^{\mu},\\
\hat{P}^{\mu} &=&
\frac{1}{1+\frac{\ell^2}{2}\hat{\textbf{p}}^2}\hat{p}^{\mu},
\end{eqnarray}
where $\hat{x}^{\mu}$ and $\hat{p}^{\mu}$ are the conventional
position and momentum operators which satisfy the conventional
covariant Heisenberg algebra $(i.e.,
\left[\hat{x}^{\mu},\hat{p}^{\nu}\right]=-i\hbar \eta^{\mu\nu} \text{and}
\left[\hat{x}^{\mu},\hat{x}^{\nu}\right]=
\left[\hat{p}^{\mu},\hat{p}^{\nu}\right]=0)$. In Eq. (10)
$\hat{\textbf{p}}^2= -\hbar^{2} \square$. Equations (9) and (10)
show that in order to reformulate quantum field theory from the
viewpoint of covariant  Kempf-Mangano algebra, the conventional
position and derivative operators $(\hat{x}^{\mu},\partial_\mu)$
must be replaced as follows:
\begin{eqnarray}
 \hat{x}^{\mu}\longrightarrow \hat{X}^{\mu} &=& \hat{x}^{\mu}, \\
 \partial_{\mu}\longrightarrow \nabla _{\mu}
 &:=&\frac{1}{1-\frac{(\hbar\ell)^2}{2}\square}\partial_{\mu}.
\end{eqnarray}
It is important to note that in the limit of $\hbar\ell\rightarrow
0$, the generalized derivative operator $\nabla _{\mu}$ in Eq. (12)
becomes the conventional derivative operator, i.e.,
$\lim_{\hbar\ell\to 0}\nabla _{\mu}=\partial_{\mu} $. In the next
section, we will introduce a Lorentz-invariant infinite derivative
scalar field theory in the framework of the covariant Kempf-Mangano
algebra.

\section{Lagrangian Formulation of an Infinite Derivative Scalar Field Theory Based on the Covariant Kempf-Mangano Algebra }

The Lagrangian density for a real scalar field in a
$(D+1)$-dimensional flat space-time can be written as follows [18]:
\begin{equation}
{\cal
L}(\phi,\partial_{\mu}\phi)=\frac{1}{2}\left(\partial_{\mu}\phi(x)\right)\left(\partial^{\mu}\phi(x)\right)-\frac{1}{2}\left(\frac{mc}{\hbar}\right)^{2}\phi^{2}(x).
\end{equation}
Using Eq. (11) together with the transformation rule for a scalar
field, we obtain
\begin{equation}
\phi(x)\longrightarrow\Phi(X)=\phi(x).
\end{equation}
If we use Eqs. (12)-(14), we will get the generalized  Lagrangian
density for a real scalar field as follows:
\begin{eqnarray}
\nonumber {\cal L} &=&
\frac{1}{2}\left(\nabla_{\mu}\Phi(X)\right)\left(\nabla^{\mu}\Phi(X)\right)-\frac{1}{2}\left(\frac{mc}{\hbar}\right)^{2}\Phi^{2}(X)
\\
 &=& \frac{1}{2}\left(\frac{1}{1-\frac{(\hbar
\ell)^2}{2}\square}\partial_{\mu}\phi(x)\right)
\left(\frac{1}{1-\frac{(\hbar \ell)^2}{2}\square}\partial
^{\mu}\phi(x)\right)
-\frac{1}{2}\left(\frac{mc}{\hbar}\right)^{2}\phi^{2}(x).
\end{eqnarray}
Now, let us consider a classical scalar field theory whose action
functional is given by [19]
\begin{equation}
S[\phi]= \int_{\mathbb{R}^{1,D}} d^D x \ dt \ {\cal
L}(\phi,\partial_{\nu_1}\phi,\partial_{\nu_1}\partial_{\nu_2}\phi,\partial_{\nu_1}\partial_{\nu_2}\partial_{\nu_3}\phi,\cdots).
\end{equation}
The variation of (16) with respect to $\phi$ leads to the following
generalized Euler-Lagrange equation [19]
\begin{eqnarray}
\frac{\partial{\cal L}}{\partial\phi}-\left(\frac{\partial{\cal
L}}{\partial\phi_{\mu_1}}\right)_{\mu_1}+\left(\frac{\partial{\cal
L}}{\partial\phi _{\mu_1\mu_2}}\right)_{\mu_1\mu_2}-\cdots +
(-1)^{k}\left(\frac{\partial{\cal L}}{\partial\phi_{\mu_1\mu_2\cdots
\mu_k}}\right)_{\mu_1\mu_2\cdots\mu_k}+\cdots=0,
\end{eqnarray}
where
\begin{eqnarray}
\phi_{\mu_1\mu_2\cdots\mu_k} :=
\partial_{\mu_1}\partial_{\mu_2}\cdots\partial_{\mu _k}\phi,
\end{eqnarray}
\begin{eqnarray}
\frac{\partial\phi_{\mu_1\mu_2\cdots\mu_k}}{\partial\phi_{\nu_1\nu_2\cdots\nu_k}}=
\delta_{\mu_1}^{\nu_1} \delta_{\mu_2}^{\nu_2} \cdots
\delta_{\mu_k}^{\nu_k}.
\end{eqnarray}
If we substitute (15) into (17), we will obtain the following
generalized Klein-Gordon wave equation
\begin{equation}
\frac{1}{\left(1-\frac{(\hbar\ell)^2}{2}\square\right)^2} \square
\phi(x) + \left(\frac{mc}{\hbar}\right)^2 \phi(x)=0.
\end{equation}
Note that in the low-energy limit $(\hbar\ell\rightarrow0)$, the
generalized Klein-Gordon equation (20) becomes the conventional
Klein-Gordon equation, i.e.,
\begin{equation}
\square\phi(x)+\left(\frac{mc}{\hbar}\right)^2
\phi(x)+\underbrace{(\hbar\ell)^2 \square\square
\phi(x)+\frac{3}{4}(\hbar\ell)^4 \square\square\square\phi(x)+{\cal
O}\left((\hbar\ell)^6 \right)}_{high-energy \;(short-distance) \;
corrections}=0.
\end{equation}
The generalized Klein-Gordon equation (20) has a plane-wave solution
as follows:
\begin{equation}
\phi(x)=A \; e^{-\frac{i}{\hbar}\mathbf{p}.\mathbf{x}},
\end{equation}
where $A$ is the amplitude of the scalar field. After inserting Eq.
(22) into Eq. (20), we obtain the following generalized dispersion
relation:
\begin{equation}
\frac{\mathbf{p}^2}{(1+\frac{\ell^2}{2}\mathbf{p}^2)^2}=m^2 c^2.
\end{equation}
Equation (23) leads to the following generalized energy-momentum
relations:
\begin{eqnarray}
E_{+}^{2}(\ell)&=& m_{+}^{2}(\ell)c^4+ c^2 \vec{\mathbf{p}}^2,
\end{eqnarray}
\begin{eqnarray}
E_{-}^{2}(\ell)&=& m_{-}^{2}(\ell)c^4+ c^2 \vec{\mathbf{p}}^2,
\end{eqnarray}
where the effective masses $m_{+}(\ell)$ and $m_{-}(\ell)$ have the
following definitions
\begin{eqnarray}
m_{+}(\ell)&:=& \frac{1+\sqrt{1-2 m^2 c^2 \ell^2}}{m c^2 \ell^2},
\end{eqnarray}
\begin{eqnarray}
m_{-}(\ell)&:=& \frac{1-\sqrt{1-2 m^2 c^2 \ell^2}}{m c^2 \ell^2}.
\end{eqnarray}
In order to obtain the real values for the effective masses
$m_{\pm}(\ell)$ in Eqs. (26) and (27) the parameter $\ell$ must
satisfy the following inequality
\begin{equation}
\ell \leqslant\frac{1}{\sqrt{2}} \frac{1}{mc}.
\end{equation}
It must be emphasized that the parameter $\hbar\ell$ which has a
dimension of $[length]$ defines a characteristic length scale
$\delta := \hbar\ell$ in our model. Equation (28) shows that the
upper limit of $\delta$ is
\begin{equation}
\delta_{max} = \frac{1}{\sqrt{2}} \frac{\hbar}{mc}.
\end{equation}
If we expand the effective masses (26) and (27) around $\delta=0$,
we will obtain the following low-energy expressions for
$m_{\pm}(\ell)$
\begin{eqnarray}
m_{+}(\ell)&=& \frac{2}{m c^2 \ell^2}-m + {\cal O}(\ell^2),
\end{eqnarray}
\begin{eqnarray}
m_{-}(\ell)&=& m+\frac{1}{2}m^3 c^2 \ell^2 +{\cal O}(\ell^4).
\end{eqnarray}
Therefore, the low-energy limit of our model describes two
particles, one with the usual mass $m$ and the other a ghostlike
particle of mass $\frac{2}{mc^2 \ell^2}$.\footnote{The appearance of
these ghostlike particles in the theory of fourth-order derivative
wave equations such as $\square \phi(x) + (\frac{mc}{\hbar})^2
\phi(x) +\frac{1}{\Lambda ^2}\square\square\phi(x)=0$ ($\Lambda$ is
a regulator (cutoff)) is a well-known problem in higher-derivative
quantum field theories
[16,20].}\\

\section{Relationship between the Low-Energy Behavior of the Model for a Spatially Homogeneous Field Configuration and the Pais-Uhlenbeck Oscillator}
In this section, we want to study the low-energy behavior of the
infinite derivative scalar field theory which was introduced in the
previous section for a spatially homogeneous field configuration.
The action functional (16) for the generalized Lagrangian density
(15) is
\begin{equation}
S_{\delta}[\phi]= \int_{\mathbb{R}^{1,D}} d^D x \ dt \ \left[{1\over
2}\left(\frac{1}{1-\frac{\delta^2}{2}\square}\partial_{\mu}\phi(x)\right)\left(\frac{1}{1-\frac{\delta^2}{2}\square}\partial^{\mu}\phi(x)\right)-{1\over
2}\left(\frac{mc}{\hbar}\right)^2\phi^2(x)\right].
\end{equation}
For a spatially homogeneous real scalar field $\phi(t)$ Eq. (32)
becomes
\begin{equation}
S_{\delta}[\phi]= \underbrace{\int_{\mathbb{R}^{D}} d^D
x}_{V_D}\int_{\mathbb{R}}\ dt \ \left[{1\over
2c^2}\left(\frac{1}{1-\frac{1}{2}\left(\frac{\delta}{c}\right)^2\frac{d^2}{dt^2}}\dot{\phi}(t)\right)\left(\frac{1}{1-\frac{1}{2}\left(\frac{\delta}{c}\right)^2\frac{d^2}{dt^2}}\dot{\phi}(t)\right)-{1\over
2}\left(\frac{mc}{\hbar}\right)^2\phi^2(t)\right],
\end{equation}
where $V_D=\int_{\mathbb{R}^{D}} d^D x$ is the volume of the spatial
part of the space-time and dot denotes derivative with respect to
$t$.\\
Using the field redefinition $\psi(t) :=
\frac{\sqrt{V_D}}{c}\phi(t)$ Eq. (33) can be rewritten as follows:
\begin{equation}
S_{\delta}[\psi]=\int_{\mathbb{R}}\ dt \ \left[{1\over
2}\left(\frac{1}{1-\frac{1}{2}\left(\frac{\delta}{c}\right)^2\frac{d^2}{dt^2}}\dot{\psi}(t)\right)\left(\frac{1}{1-\frac{1}{2}\left(\frac{\delta}{c}\right)^2\frac{d^2}{dt^2}}\dot{\psi}(t)\right)-{1\over
2}\left(\frac{mc^2}{\hbar}\right)^2\psi^{2}(t)\right].
\end{equation}
The action functional $S_{\delta}[\psi]$ in Eq. (34) has the
following low-energy expansion
\begin{eqnarray}
\nonumber S_{\delta}[\psi]&=& \sum_{n=0}^{\infty} \delta^{2n}
S_{n}[\psi]\\
\nonumber &=& \int_{\mathbb{R}}\ dt \ \left[{1\over
2}\left(\dot{\psi}^{2}(t)+\left(\frac{\delta}{c}\right)^{2}\dot{\psi}(t)\
\dddot{\psi}(t)\right)-{1\over
2}\left(\frac{mc^2}{\hbar}\right)^2\psi^{2}(t)\right]+{\cal
O}(\delta^4)\\
&=& \int_{\mathbb{R}}\ dt\ \left[{1\over
2}\left(\dot{\psi}^{2}(t)-\left(\frac{\delta}{c}\right)^{2}
\ddot{\psi}^{2}(t)\right)-{1\over
2}\left(\frac{mc^2}{\hbar}\right)^2\psi^{2}(t)\right]+\int_{\mathbb{R}}\
dt \frac{d\Omega(t)}{dt}+{\cal O}(\delta^4),
\end{eqnarray}
where
\begin{eqnarray}
\Omega(t):= {1\over 2}(\frac{\delta}{c})^{2} \dot{\psi}(t)
\ddot{\psi}(t).
\end{eqnarray}
If we neglect terms of order $\delta^4$ and higher in Eq. (35) and
dropping out the boundary term $\int_{\mathbb{R}}\ dt
\frac{d\Omega(t)}{dt}$, we will find
\begin{eqnarray}
S_{\delta}[\psi]=\int_{\mathbb{R}}\ dt \ \left[{1\over
2}\dot{\psi}^{2}(t)-{1\over
2}\left(\frac{\delta}{c}\right)^{2}\ddot{\psi}^{2}(t)-{1\over
2}\left(\frac{mc^2}{\hbar}\right)^2\psi^{2}(t)\right].
\end{eqnarray}
Straightforward but tedious calculations show that
$S_{\delta}[\psi]$ in Eq. (37) can be written as follows:
\begin{eqnarray}
S_{\delta}[\psi]=-{1\over 2}\left(\frac{\delta}{c}\right)^{2}
\int_{\mathbb{R}}\ dt \
\left[\ddot{\psi}^{2}(t)-(\omega_{+}^{2}+\omega_{-}^{2})\dot{\psi}^{2}(t)+\omega_{+}^{2}\omega_{-}^{2}\psi^{2}(t)\right],
\end{eqnarray}
where the effective frequencies $\omega_{\pm}$ are defined as
follows:
\begin{eqnarray}
\omega_{+}:= {c\over
2\delta}\left[\sqrt{1+{2mc\delta\over\hbar}}+\sqrt{1-{2mc\delta\over\hbar}}\right],
\end{eqnarray}
\begin{eqnarray}
\omega_{-}:= {c\over
2\delta}\left[\sqrt{1+{2mc\delta\over\hbar}}-\sqrt{1-{2mc\delta\over\hbar}}\right].
\end{eqnarray}
The action functional (38) is a well-known model in the theory of
higher-order time derivative models and is called the Pais-Uhlenbeck
(PU) oscillator [4,21-25].\footnote{The Pais-Uhlenbeck oscillator
describes a one-dimensional harmonic oscillator coupled to a
higher-order time derivative term whose action functional is
\begin{eqnarray*}
S_{PU}={\gamma\over 2} \int_{\mathbb{R}}\ dt \
\left[\ddot{\psi}^{2}(t)-(\omega_{+}^{2}+\omega_{-}^{2})\dot{\psi}^{2}(t)+\omega_{+}^{2}\omega_{-}^{2}\psi^{2}(t)\right],
\end{eqnarray*}
where $\gamma$ is an arbitrary parameter [21]. This model has a wide
applications in several areas of theoretical physics [21-25].}\\
Therefore, in the low-energy limit $(\delta\rightarrow0)$ our model behaves like a Pais-Uhlenbeck oscillator for a spatially
homogeneous field configuration $\phi(t,\vec{\textbf{x}})=\phi(t)$.
In order to obtain the real values for the effective frequencies
$\omega_{\pm}$ in Eqs. (39) and (40) the characteristic length scale
$\delta$ must satisfy the following condition
\begin{equation}
\delta^{PU} \leqslant\frac{1}{2} \frac{\hbar}{mc}.
\end{equation}
According to Eq. (41) the upper limit of $\delta^{PU}$ is
\begin{equation}
\delta_{max}^{PU}=\frac{1}{2} \frac{\hbar}{mc}.
\end{equation}
Equations (29) and (42) show that the upper limit of the
characteristic length scale $\delta$ in this work is proportional to $\frac{\hbar}{mc}$, i.e.,
\begin{equation}
\delta_{max}\sim \frac{\hbar}{mc}.
\end{equation}

\section{Summary and Conclusions}
More than 70 years ago the American physicist H. S. Snyder introduced a one-parameter extension of the covariant Heisenberg algebra
in a four-dimensional space-time in order to remove the infinities which appear in quantum field theories [26].
In 2006 a two-parameter extension of the covariant Heisenberg algebra in a $(D+1)$-dimensional Minkowski space-time was presented by Quesne and Tkachuk [15].
The Quesne-Tkachuk algebra contains the Snyder algebra as a subalgebra [15]. In addition, the reformulation of Maxwell equations and Dirac equation from the
viewpoint of the Quesne-Tkachuk algebra have been studied for the first time in Ref. [27].\\
In 2017, G. P. de Brito and his co-workers introduced a modification
of the Quesne-Tkachuk algebra [14]. This modified algebra is a
covariant generalization of the Kempf-Mangano algebra in a
$(D+1)$-dimensional Minkowski space-time. In this work, by using the methods of Ref. [27], after
Lagrangian formulation of an infinite derivative scalar field theory
in the framework of the covariant Kempf-Mangano algebra, it was
shown that the infinite derivative field equation (20) describes two
particles with the effective masses $ m_{\pm}(\ell)=
\frac{1\pm\sqrt{1-2 m^2 c^2 \ell^2}}{m c^2 \ell^2}$. We showed that
in the low-energy (large-distance) limit the infinite derivative
scalar field theory in Eq. (32) for a spatially homogeneous field
configuration $\phi(t,\vec{\textbf{x}})=\phi(t)$ behaves like a
Pais-Uhlenbeck oscillator with the effective frequencies
$\omega_{\pm}= {c\over
2\delta}\left[\sqrt{1+{2mc\delta\over\hbar}}\pm\sqrt{1-{2mc\delta\over\hbar}}\right]$,
where $\delta=\hbar\ell$ is the characteristic length scale in this
paper. Our calculations in Sections 3 and 4 show that the upper
limit of $\delta$ must be proportional to $\frac{\hbar}{mc}$ (see
Eq. (43)).\\Now, let us evaluate the numerical value of
$\delta_{max}$ in Eq. (43).\\In nuclear and low-energy particle
physics a real scalar field theory describes a neutral $\pi^0$ meson
[28]. The mass of the $\pi^0$ meson is [18]
\begin{equation}
m_{\pi^0}=134.977 \, \, MeV/c^2.
\end{equation}
Inserting (44) into (43), we find
\begin{equation}
\delta_{max}\sim 10^{-15} m.
\end{equation}
It should be noted that the numerical value of $\delta_{max}$ in Eq.
(45) is near to the nuclear scale (see page 174 in Ref. [29]), i.e.,
\begin{equation}
\delta_{max}\sim \delta_{nuclear\ scale}\sim 10^{-15}\, m.
\end{equation}
The above estimations show that in the low-energy limit, the conventional real scalar
field theory in Eq. (13) is recovered, while in the high-energy limit the real
Klein-Gordon theory in Eq. (13) must be replaced by Eq. (15), i.e.,
\begin{equation}
{\cal L}=\begin{cases}
\frac{1}{2}\left(\partial_{\mu}\phi(x)\right)\left(\partial^{\mu}\phi(x)\right)-\frac{1}{2}\left(\frac{mc}{\hbar}\right)^{2}\phi^{2}(x)
& \text{(low-energy regime)},\\
\frac{1}{2}\left(\partial_{\mu}\phi(x)\right)\left(\partial^{\mu}\phi(x)\right)-\frac{1}{2}\left(\frac{mc}{\hbar}\right)^{2}\phi^{2}(x)\\-{1\over
2} \delta^2(\square\phi(x))(\square\phi(x))+boundary\;term +{\cal
O}(\delta^4) & \text{(high-energy regime)}.
\end{cases}
\end{equation}
The action functional (32) in the presence of an external current
$J(x)$ is
\begin{equation}
S_{\delta}[\phi]={1\over c}\int_{\mathbb{R}^{1,D}} d^{D+1} x
\left[{1\over
2}\left(\frac{1}{1-\frac{\delta^2}{2}\square}\partial_{\mu}\phi(x)\right)\left(\frac{1}{1-\frac{\delta^2}{2}\square}\partial^{\mu}\phi(x)\right)-{1\over
2}\left(\frac{mc}{\hbar}\right)^2\phi^2(x)+J(x) \phi(x)\right].
\end{equation}
The action functional (48) can be rewritten as follows:
\begin{equation}
S_{\delta}[\phi]={1\over c}\int_{\mathbb{R}^{1,D}} d^{D+1} x
\left[-{1\over
2}\phi(x)\left(\nabla_{\mu}\nabla^{\mu}+\left(\frac{mc}{\hbar}\right)^2\right)\phi(x)+\partial_{\mu}\Upsilon^{\mu}+J(x) \phi(x)\right],
\end{equation}
where $\nabla_\mu$ has been defined in Eq. (12) and $\Upsilon^{\mu}$ has the following definition
\begin{equation}
\Upsilon^{\mu}:=\frac{1}{2}\phi\partial^{\mu}\phi+\frac{\delta^{2}}{2}\phi\square\partial^{\mu}\phi+\frac{\delta^{4}}{8}\left(3\phi\square^{2}\partial^{\mu}\phi-\partial^{\mu}\phi
\square^{2}\phi+\square\phi\square\partial^{\mu}\phi\right)+{\cal O}(\delta^6).
\end{equation}
After dropping out the boundary term ${1\over c}\int_{\mathbb{R}^{1,D}} d^{D+1}x \ \partial_{\mu}\Upsilon^{\mu}$ in 
(49), we will find
\begin{equation}
S_{\delta}[\phi]={1\over c}\int_{\mathbb{R}^{1,D}} d^{D+1} x
\left[-{1\over
2}\phi(x)\left(\nabla_{\mu}\nabla^{\mu}+\left(\frac{mc}{\hbar}\right)^2\right)\phi(x)+J(x)
\phi(x)\right].
\end{equation}
A comparison between the action functionals (2) and (51) shows that
for $V\left(\phi(x)\right)={1\over
2}\left(\frac{mc}{\hbar}\right)^2\phi^2(x)$ and
$F(\square_x)=-\frac{\square_x}{(1-\frac{\delta^2}{2}\square_x)^2}$
there is an equivalence between a non-local real scalar field theory
and an infinite derivative real scalar field theory in the framework
of the covariant Kempf-Mangano algebra.\footnote{Note that the form
factor $F(z)=-\frac{z}{(1-\frac{\delta^2}{2}z)^2}$ is not an entire
function. It must be emphasized that there are many examples in the
literatures about non-local quantum field theory in which the form
factor $F(z)$ is not an entire function (see Refs. [10,30,31] for more details).}



\section*{Acknowledgments}
We would like to thank the referee for his/her careful reading and
constructive comments.

\end{document}